\documentclass[aps, pra, 10pt, twocolumn, letterpaper, superscriptaddress, nofootinbib]{revtex4}

\usepackage{amsmath}
\usepackage{amssymb}
\usepackage{mathrsfs}
\usepackage{xspace}
\usepackage{braket}
\usepackage{bbold}
\usepackage{graphicx}
\usepackage{nicefrac}
\usepackage{color}
\usepackage{ulem}

\usepackage{ifpdf}
\ifpdf
\fi

\newcommand{\myRef}[1]{Ref.~\cite{#1}}
\ifx\Ref\undefined
\newcommand{\Ref}[1]{\myRef{#1}}
\else
\renewcommand{\Ref}[1]{\myRef{#1}}
\fi
\newcommand{\eq}[1]{(\ref{#1})}
\newcommand{\Eq}[1]{Eq.~(\ref{#1})}
\newcommand{\Eqs}[1]{Eqs.~(\ref{#1})}

\newcommand{\Sec}[1]{Sec.~\ref{#1}}
\newcommand{\Refs}[1]{Refs.~\cite{#1}}
\newcommand{\App}[1]{Appendix~\ref{#1}}
\newcommand{\Apps}[1]{Appendices~\ref{#1}}

\newcommand{\ie}{{i.e.,\/}\xspace}

\newcommand{\pd}{\partial}
\newcommand{\del}{\nabla}
\newcommand{\mc}[1]{\mathcal{#1}}

\newcommand{\msf}[1]{\mathsf{#1}}
\newcommand{\mcu}[1]{\mathscr{#1}}

\newcommand{\oper}[1]{\smash{\widehat{#1}}}
\newcommand{\boper}[1]{\oper{\boldsymbol{#1}}}
\renewcommand{\vec}[1]{{\boldsymbol{#1}}}
\newcommand{\matr}[1]{{\vec{#1}}}

\newcommand{\ii}{\mathrm{i}}
\newcommand{\dd}{\mathrm{d}}

\newcommand{\total}[1]{\msf{#1}}
\newcommand{\invh}{\psi}
\newcommand{\ninvh}{\phi}
\newcommand{\mT}{V}
\newcommand{\lie}{\mathrm{\text{\pounds}}}
\newcommand{\spatialk}{\msf{k}}

\allowdisplaybreaks

\begin{document}

\title{Gauge invariants of linearized gravity with a general background metric}

\author{Deepen Garg}
\affiliation{Department of Astrophysical Sciences, Princeton University, Princeton, New Jersey 08544, USA}
\author{I. Y. Dodin}
\affiliation{Department of Astrophysical Sciences, Princeton University, Princeton, New Jersey 08544, USA}
\affiliation{Princeton Plasma Physics Laboratory, Princeton, NJ 08543, USA}

\date{\today}

\begin{abstract}
In linearized gravity with distributed matter, the background metric has no generic symmetries, and decomposition of the metric perturbation into global normal modes is generally impractical. This complicates the identification of the gauge-invariant part of the perturbation, which is a concern, for example, in the theory of dispersive gravitational waves whose energy--momentum must be gauge-invariant. Here, we propose how to identify the gauge-invariant part of the metric perturbation and the six independent gauge invariants \textit{per~se} for an arbitrary background metric. For the Minkowski background, the operator that projects the metric perturbation on the invariant subspace is proportional to the well-known dispersion operator of linear gravitational waves in~vacuum. For a general background, this operator  is expressed in terms of the Green's operator of the vacuum wave equation. If the background is smooth,  it can be found asymptotically using the inverse scale of the background metric as a small parameter.
\end{abstract}

\maketitle

\section{Introduction}
\label{sec:intro}

In many problems related to gravity, the complicated structure of the Einstein field equations necessitates a perturbative approach within which the spacetime metric is split into a background metric and a small perturbation, and the equations are often linearized in the perturbation metric \cite{book:carroll, book:mukhanov, ref:flanagan05}. While this approach allows for a tractable answer for many interesting phenomena such as gravitational waves (GWs) and Jeans theory \cite[Chap.~7]{book:mukhanov}, it also introduces a gauge freedom that has to be dealt with. To be specific, let us consider the background metric to be $\smash{g_{\alpha\beta}} = \mc{O}(1)$ and the perturbation metric to be $\smash{h_{\alpha\beta}} = \mc{O}(a)$, where $a \ll 1$ is a small parameter. A coordinate transformation $\smash{x^\mu \to x'^\mu = x^\mu + \xi^\mu}$, with $\smash{\xi^\mu} = \mc{O}(a)$, induces a metric transformation $\smash{\total{g}_{\alpha\beta}} \to \smash{\total{g}'_{\alpha\beta}} = \total{g}_{\alpha\beta} - \smash{\lie_\xi \total{g}_{\alpha\beta}} + \mc{O}(a^2)$, where $\smash{\lie_\xi}$ is the Lie derivative along the vector field~$\smash{\xi^\mu}$~\cite{book:carroll} and $\smash{\total{g}_{\alpha\beta}}$ is the total spacetime metric. Within linearized gravity, where $\smash{\mc{O}(a^2)}$ corrections are neglected and the background is $a$-independent by definition, this implies $\smash{g_{\alpha\beta}} \to \smash{g'_{\alpha\beta}} = \smash{g_{\alpha\beta}}$ and $\smash{h_{\alpha\beta}} \to \smash{h'_{\alpha\beta}} = \smash{h_{\alpha\beta}} - \smash{\lie_\xi g_{\alpha\beta}}$. If $\smash{h_{\alpha\beta}}$ is treated as a tensor field on the unperturbed spacetime, so its indices are manipulated using $\smash{g_{\alpha\beta}}$ as the metric, one also has
\begin{gather}
h^{\alpha\beta} \to h'^{\alpha\beta} = h^{\alpha\beta} + \lie_\xi g^{\alpha\beta},
\label{eq:gauge}
\end{gather}
and as a reminder,
\begin{gather}
\lie_\xi g^{\alpha\beta} = -\del^\alpha \xi^\beta - \del^\beta \xi^\alpha \equiv -2\del^{(\alpha} \xi^{\beta)}.
\label{eq:lie}
\end{gather}
The transformation \eq{eq:gauge} can be viewed as a gauge transformation (with $\xi^\mu$ being the gauge field) and, by general covariance, cannot have measurable effects. Thus, the physical, gauge-invariant, part of $\smash{h^{\alpha\beta}}$ is defined only up to the Lie derivative of $g^{\alpha\beta}$ along an arbitrary vector field, which is encoded by four functions (in a four-dimensional spacetime). Because the symmetric tensor $\smash{h^{\alpha\beta}}$ is encoded by ten functions, this leaves room for six gauge-invariant degrees of freedom.

The nonphysical components of the perturbation metric obscure physical phenomena with coordinate artifacts, making it difficult to distinguish what is real and what is not. Thus, it is important to be able to indentify the gauge-invariant degrees of freedom for a given metric perturbation and to represent the reduced equations of perturbation gravity in a gauge-invariant form. This problem has attracted considerable interest in many different contexts. In cosmological settings, the background can often be fixed to be Friedmann--Lema\^itre--Robertson--Walker metric, which then can be analyzed using Bardeen's formalism \cite{ref:bardeen80, ref:malik09, ref:malik13, book:mukhanov, ref:fewster13} or other methods, for example, using geodesic lightcone coordinates \cite{ref:fanizza21, ref:frob22}. Bardeen's formalism has also been extended to second and higher-order perturbations \cite{ref:nakamura07c, tex:nakamura19, ref:bruni97, ref:luca20} that are relevant for GWs produced in the early Universe, and also specifically for inflationary cosmologies \cite{ref:frob17, ref:frob18a}. One can take the flat spacetime limit of Bardeen's formalism to derive the gauge-invariant degrees of freedom for the Minkowski background \cite{ref:flanagan05, ref:moretti19, tex:higuchi13}. Similarly, the gauge-invariant perturbations have been studied for the Schwarzschild and Kerr background as well \cite{ref:thompson17, ref:aksteiner21, tex:nakamura21}.

While the analysis of isotropic backgrounds suffices for many settings, problems that involve GW--matter coupling \cite{my:gwponder, ref:baym17, ref:bamba18, ref:asenjo20, ref:barta18, ref:chesters73, ref:asseo76, ref:macedo83, ref:flauger18, ref:servin01, ref:moortgat03, ref:forsberg10a, ref:isliker06, ref:duez05a, ref:mendonca02b} require a more general analysis. Usually, this coupling is studied by ignoring the backreation of matter on metric oscillations \cite{ref:isliker06, ref:brodin00, ref:brodin10b, ref:brodin00b, ref:brodin01, ref:brodin05}, because the interaction of GWs with cold collisionless matter is weak \cite{ref:flauger18}. However, a more systematic theory is required to accommodate, for example, thermal effects \cite{ref:kumar19}, alternative GW polarizations \cite{ref:moretti20}, and fluid viscosity \cite{ref:madore73}. In particular, adequately describing linear transformations of GW modes in inhomogeneous matter (known as mode conversion for general waves \cite{book:tracy}) requires that all gravitational perturbations be treated on the same footing and the GW polarization be derived rather than assumed \textit{a~priori} \cite{my:gwinvar}. 

Ensuring the gauge invariance of the wave equation reported in \Ref{my:gwinvar} requires identification of the gauge-invariant variables for a general background metric. This has been studied, for example, using the Arnowitt--Deser--Misner (ADM) decomposition \cite{tex:arnowitt} for the background metric with a focus on application to the higher-order perturbations \cite{ref:nakamura13} (see also \Ref{tex:khavkine19}). However, fundamental theory of general dispersive GWs must be covariant, and also it cannot rely on symmetry considerations \cite{ref:asseo76} or normal-mode decomposition \cite{tex:arnowitt} that are commonly used for vacuum GWs. As known from plasma-wave theory \cite{book:stix}, which deals with similar issues for electromagnetic waves, these approaches become impractical once wave--matter coupling is introduced. (In particular, the wave polarization changes continuously as a function of the matter parameters \cite{my:gwinvar}, and the number of normal modes in the presence of matter is generally infinite \cite{ref:vankampen55, my:nonloc}.) Thus, formulating GW theory in a covariant gauge-invariant form remains an important problem to solve.

Here, we explicitly identify the invariant part of a metric perturbation on a general background metric within linearized gravity. We start by showing that any metric perturbation $\smash{h^{\alpha\beta}}$ can be uniquely decomposed as 
\begin{gather}
\label{eq:decomp}
h^{\alpha\beta} 
= \oper{\Pi}^{\alpha\beta}_{\rm inv} {}_{\gamma\delta} h^{\gamma\delta} 
+ \oper{\Pi}^{\alpha\beta}_{\rm g} {}_{\gamma\delta} h^{\gamma\delta},
\end{gather}
where the operators $\smash{\oper{\Pi}^{\alpha\beta}_{\rm inv} {}_{\gamma\delta}}$ and $\smash{\oper{\Pi}^{\alpha\beta}_{\rm g} {}_{\gamma\delta}}$ are projectors that satisfy
\begin{subequations}\label{eq:props}
\begin{gather}
\oper{\Pi}^{\alpha\beta}_{\rm inv} {}_{\gamma\delta}
+ \oper{\Pi}^{\alpha\beta}_{\rm g} {}_{\gamma\delta}
= \delta^\alpha_{(\gamma}\delta^\beta_{\delta)},
\label{eq:PiPi}
\\
\oper{\Pi}^{\alpha\beta}_{\rm inv} {}_{\gamma\delta}
\oper{\Pi}^{\gamma\delta}_{\rm inv} {}_{\lambda\varepsilon} 
= \oper{\Pi}^{\alpha\beta}_{\rm inv} {}_{\lambda\varepsilon},
\label{eq:proj1}
\\
\oper{\Pi}^{\alpha\beta}_{\rm g} {}_{\gamma\delta}
\oper{\Pi}^{\gamma\delta}_{\rm g} {}_{\lambda\varepsilon} 
= \oper{\Pi}^{\alpha\beta}_{\rm g} {}_{\lambda\varepsilon},
\label{eq:proj2}
\\
\oper{\Pi}^{\alpha\beta}_{\rm inv} {}_{\gamma\delta}
\oper{\Pi}^{\gamma\delta}_{\rm g} {}_{\lambda\varepsilon} 
=
\oper{\Pi}^{\alpha\beta}_{\rm g} {}_{\gamma\delta}
\oper{\Pi}^{\gamma\delta}_{\rm inv} {}_{\lambda\varepsilon} 
= 0,
\label{eq:proj0}
\\
\oper{\Pi}^{\alpha\beta}_{\rm inv} {}_{\gamma\delta} \lie_u g^{\gamma\delta} = 0,
\label{eq:PiLie}
\\
\oper{\Pi}^{\alpha\beta}_{\rm g} {}_{\gamma\delta} \lie_u g^{\gamma\delta} 
= \lie_u g^{\alpha\beta},
\label{eq:PiLie2}
\\
\oper{\Pi}^{\alpha\beta}_{{\rm inv},{\rm g}}{}_{\gamma\delta}
= \oper{\Pi}^{\alpha\beta}_{{\rm inv},{\rm g}}{}_{\delta\gamma}
= \oper{\Pi}^{\beta\alpha}_{{\rm inv},{\rm g}}{}_{\gamma\delta}.
\label{eq:Pisym}
\end{gather}
\end{subequations}
(Parentheses in indices denote symmetrization, as usual, and $u^\mu$ is any vector field.) In \Sec{sec:decomp}, we present a method for how to calculate the operators $\smash{\oper{\Pi}^{\alpha\beta}_{\rm inv} {}_{\gamma\delta}}$ and $\smash{\oper{\Pi}^{\alpha\beta}_{\rm g} {}_{\gamma\delta}}$ for general $g_{\alpha\beta}$. We also show that the gauge-invariant part of a metric perturbation $\smash{h^{\alpha\beta}}$ can be calculated as $\smash{\oper{\Pi}^{\alpha\beta}_{\rm inv} {}_{\gamma\delta} h^{\gamma\delta}}$, while $\smash{\oper{\Pi}^{\alpha\beta}_{\rm g} {}_{\gamma\delta} h^{\gamma\delta}}$ is the gauge-dependent part representable as $\smash{\lie_\zeta g^{\alpha\beta}}$, where $\zeta^\mu$ is a vector field linear in $\smash{h^{\alpha\beta}}$. In \Sec{sec:eucl}, we illustrate the application of our results to the Minkowski background as an explicitly solvable problem that allows benchmarking our theory against known results.\footnote{For GW modes of certain types \cite{my:gwinvar}, the Minkowski-background model can also be relevant for studies of GW--matter coupling.} We derive the six gauge-invariant components of $\smash{h^{\alpha\beta}}$ and show the agreement with the commonly known results. In addition, we show that $\smash{\oper{\Pi}^{\alpha\beta}_{\rm inv} {}_{\gamma\delta}}$ is proportional to the dispersion operator of linear GWs in Minkowski vacuum. In \Sec{sec:conclusion}, we summarize our main results. Other auxiliary calculations are presented in \Apps{app:Xi} through \ref{app:del2normal}.

\section{Expressions for the operators $\boldsymbol{\smash{\oper{\Pi}^{\alpha\beta}_{\rm inv} {}_{\gamma\delta}}}$ and $\boldsymbol{\smash{\oper{\Pi}^{\alpha\beta}_{\rm g} {}_{\gamma\delta}}}$}
\label{sec:decomp}

In this section, we derive explicit expressions for the operators $\smash{\oper{\Pi}^{\alpha\beta}_{\rm inv} {}_{\gamma\delta}}$ and $\smash{\oper{\Pi}^{\alpha\beta}_{\rm g} {}_{\gamma\delta}}$ for a general $g_{\alpha\beta}$. We assume the sign convention as in \Refs{book:carroll, book:misner77}, so
\begin{gather}\label{eq:Ru}
[\del_\beta, \del^\alpha] \xi^\beta = R^\alpha{}_\beta \xi^\beta
\end{gather}
for any vector field $\xi^\alpha$, where $\smash{R^\alpha{}_\beta}$ is the Ricci tensor. We assume that the matter is localized, so that at sufficiently large spatial distances (defined, say, in the center-of-mass frame), $R^\alpha{}_\beta$ vanishes and they satisfy outgoing boundary conditions, \ie no GWs are going in through a sufficiently large spatial sphere. Specifically, since GWs are vacuum tensor modes at infinity, we assume $\pd_0 - n^i \pd_i = 0$, where $n^i$ is normal to the spatial sphere (the polarization does not have to be specified), and the Latin index $i$ represents spatial coordinates. We assume that this applies to both $h^{\alpha\beta}$ and $h'^{\alpha\beta}$ and therefore to gauge fields as well. Then, one can proceed as follows.

\subsection{Special case}
\label{sec:special}

To motivate the machinery that will be introduced in \Sec{sec:gen}, let us first discuss an auxiliary problem. Consider a vector field $\smash{u^\alpha}$ that transforms the gauge of a given metric perturbation $\smash{h^{\alpha\beta}}$ to the Lorenz gauge:
\begin{gather}
h'^{\alpha\beta} \doteq h^{\alpha\beta} - \lie_u g^{\alpha\beta}
\label{eq:hpe},
\\
\del_\beta h'^{\alpha\beta} = 0,
\label{eq:hpLg}
\end{gather}
where the symbol $\doteq$ denotes definitions. Let us assume \textit{for now} that $\smash{u^\alpha}$ is divergence-free; \ie
\begin{gather}\label{eq:xidiv}
\del_\alpha u^\alpha = 0.
\end{gather}
Then, \Eqs{eq:lie}--\eq{eq:hpLg} yield
\begin{subequations}
\begin{gather}
\oper{Q}^\alpha{}_\beta u^\beta = \del_\beta h^{\alpha\beta},
\label{eq:xih}
\\
\oper{Q}^\alpha{}_\beta \doteq - \delta_\beta^\alpha \del_\mu \del^\mu - R^\alpha{}_\beta,
\label{eq:QX}
\end{gather}
\end{subequations}
where we have used \Eq{eq:Ru}. The hyperbolic operator \(\oper{Q}^\alpha{}_\beta\) is similar to the one that appears in the driven Maxwell's equation for the Lorenz-gauge electromagnetic vector potential in vacuum \cite{my:spinhall} except for the opposite sign in front of the Ricci tensor. In the presence of matter, GWs are dispersive (vacuum waves can be considered as a limit; see \Sec{sec:eucl}), so $\oper{Q}^\alpha{}_\beta$ is generally invertible for fields of interest under the assumed boundary conditions. Then, one can introduce a unique Green's operator of \Eq{eq:xih} as
\begin{gather}
\oper{\Xi}^\alpha{}_\beta = (\oper{Q}^{-1})^\alpha{}_\beta
\label{eq:Xi}
\end{gather}
and express the solution of \Eq{eq:xih} as follows:
\begin{gather}\label{eq:xisol}
u^\alpha = \oper{\Xi}^\alpha{}_{(\gamma} \del_{\delta)} h^{\gamma\delta},
\end{gather}
where symmetrization with respect to the lower indices is added for convenience and does not affect the result. (As a side remark, the appearance of Green's operator is not unexpected here; cf.~\Ref{ref:nakamura13}, where Green's functions of elliptic operators appear in a related problem for ADM-parameterized backgrounds.) Finding $\smash{(\oper{Q}^{-1})^\alpha{}_\beta}$ is equivalent to finding waves generated by prescribed sources. Similar calculations for driven Maxwell's equation in various covariant gauges can be found in \Ref{ref:frob18b}. See also \Ref{my:ql}.

Since, $\smash{h^{\gamma\delta}}$ is assumed to be such that the solution \eq{eq:xisol} satisfies the constraint \eq{eq:xidiv}, \Eq{eq:hpLg} is satisfied by
\begin{gather}
\smash{h'^{\alpha\beta}} = \smash{\oper{\pi}^{\alpha\beta}{}_{\gamma\delta} h^{\gamma\delta}},    
\end{gather}
where we defined
\begin{gather}\label{eq:P0}
\oper{\pi}^{\alpha\beta}{}_{\gamma\delta} \doteq
\delta^\alpha_{(\gamma}\delta^\beta_{\delta)}
+ 2\del^{(\alpha} 
\oper{\Xi}^{\beta)}{}_{(\gamma} \del_{\delta)}.
\end{gather}
In combination with \Eq{eq:hpe}, these results yield that
\begin{subequations}\label{eq:Ph}
\begin{gather}
h^{\alpha\beta} = \oper{\pi}^{\alpha\beta}{}_{\gamma\delta} h^{\gamma\delta} 
+ \lie_{u} g^{\alpha\beta},\\
\lie_{u} g^{\alpha\beta} = -2\del^{(\alpha}
\oper{\Xi}^{\beta)}{}_{(\gamma} \del_{\delta)} h^{\gamma\delta},
\end{gather}
\end{subequations}
and a direct calculation shows that (\App{app:PLie})
\begin{gather}\label{eq:PLie0}
\oper{\pi}^{\alpha\beta}{}_{\gamma\delta} \lie_u g^{\gamma\delta} = 0.
\end{gather}
Equation \eq{eq:PLie0} is similar to \Eq{eq:PiLie} and makes the decomposition \eq{eq:Ph} close to \Eq{eq:decomp}, except it is constrained by \Eq{eq:xidiv}. This can be taken as a hint that $\smash{\oper{\pi}^{\alpha\beta}{}_{\gamma\delta}}$ is close to the sought $\smash{\oper{\Pi}^{\alpha\beta}_{\rm inv} {}_{\gamma\delta}}$. Hence, we approach the general case as follows.

\subsection{General case}
\label{sec:gen}

Now let us waive the Lorenz-gauge assumption \eq{eq:xidiv} and consider applying $\oper{\pi}^{\alpha\beta}{}_{\gamma\delta}$ to $\smash{\lie_u g^{\alpha\beta}}$ with a general $\smash{u^\alpha}$. In this case, \Eq{eq:hpLg} is not necessarily satisfied, but a direct calculation shows that (\App{app:PLie})\footnote{Here and further, $g_{\alpha\beta} \equiv \oper{g}_{\alpha\beta}$ and $R^\alpha{}_\beta \equiv \oper{R}^\alpha{}_\beta$ serve as multiplication operators, and the assumed notation is $\smash{\oper{A}\oper{B}f = \oper{A}(\oper{B}f)}$ for any operators $\smash{\oper{A}}$ and $\smash{\oper{B}}$ and function $f$ that they act upon. For example, $\del^\mu g_{\gamma\delta} \lie_u g^{\gamma\delta} \equiv \del^\mu [g_{\gamma\delta} (\lie_u g^{\gamma\delta})]$.}
\begin{gather}\label{eq:PLie}
\oper{\pi}^{\alpha\beta}{}_{\gamma\delta} \lie_u g^{\gamma\delta} 
=
\del^{(\alpha} 
\oper{\Xi}^{\beta)}{}_{\mu} \del^\mu g_{\gamma\delta} \lie_u g^{\gamma\delta}.
\end{gather}
Hence, the operator
\begin{gather}\label{eq:PiMain}
\oper{\Pi}^{\alpha\beta}_{\rm inv} {}_{\gamma\delta} \doteq \oper{\pi}^{\alpha\beta}{}_{\gamma\delta} -
\del^{(\alpha} \oper{\Xi}^{\beta)}{}_{\mu} \del^\mu g_{\gamma\delta}
\end{gather}
automatically satisfies \Eq{eq:PiLie}. Let us substitute \Eq{eq:P0} and rewrite this operator as follows:
\begin{subequations}
\label{eq:projectors}
\begin{gather}
\oper{\Pi}^{\alpha\beta}_{\rm inv} {}_{\gamma\delta} 
=
\delta^\alpha_{(\gamma}\delta^\beta_{\delta)}
- \oper{\Pi}^{\alpha\beta}_{\rm g} {}_{\gamma\delta},
\label{eq:PiPi2}\\
\oper{\Pi}^{\alpha\beta}_{\rm g} {}_{\gamma\delta}
\doteq
-2\del^{(\alpha} \oper{\Xi}^{\beta)}{}_{(\gamma} \del_{\delta)}
+\del^{(\alpha} \oper{\Xi}^{\beta)}{}_{\mu} \del^\mu g_{\gamma\delta}.
\label{eq:Piparallel}
\end{gather}
\end{subequations}
This satisfies \Eqs{eq:PiPi}, \eq{eq:PiLie2}, and \eq{eq:Pisym}. (The latter ensures that $\smash{\oper{\Pi}^{\alpha\beta}_{{\rm inv},{\rm g}}{}_{\gamma\delta} f^{\gamma\delta}} = 0$ for all anti-symmetric $\smash{f^{\gamma\delta}}$, which is convenient.) The property~\eq{eq:proj2} is proven by a direct calculation (\App{app:proj}). Equation \eq{eq:proj0} can be derived from \Eqs{eq:PiPi} and \eq{eq:proj2}, and the remaining property~\eq{eq:proj1} can then be obtained from \Eqs{eq:PiPi} and \eq{eq:proj0}.

Let us discuss how this intermediate result helps identify the invariant part of a metric perturbation. First, notice that
\begin{align}
\oper{\Pi}^{\alpha\beta}_{\rm g} {}_{\gamma\delta}h^{\gamma\delta}
& = -2\del^{(\alpha} \oper{\Xi}^{\beta)}{}_{(\gamma} \del_{\delta)}h^{\gamma\delta}
+\del^{(\alpha} \oper{\Xi}^{\beta)}{}_{\mu} \del^\mu g_{\gamma\delta}h^{\gamma\delta}
\notag\\ 
& = - 2\del^{(\alpha} \zeta^{\beta)} 
\notag\\
& = \lie_\zeta g^{\alpha\beta},
\end{align}
where we introduced
\begin{gather}
\zeta^\beta \doteq \oper{\Xi}^{\beta}{}_{(\gamma} \del_{\delta)}h^{\gamma\delta}
- \frac{1}{2}\,\oper{\Xi}^{\beta}{}_{\mu} \del^\mu g_{\gamma\delta}h^{\gamma\delta}.
\label{eq:zeta}
\end{gather}
Hence, \Eq{eq:decomp} can be rewritten as
\begin{subequations}
\label{eq:gaugeinvh}
\begin{gather}
h^{\alpha\beta} = \invh^{\alpha\beta} + \ninvh^{\alpha\beta} ,
\label{eq:gaugeinvh1}
\\
\invh^{\alpha\beta} \doteq \oper{\Pi}^{\alpha\beta}_{\rm inv} {}_{\gamma\delta} h^{\gamma\delta},
\label{eq:gaugeinvh2}
\\
\ninvh^{\alpha\beta} \doteq \oper{\Pi}^{\alpha\beta}_{\rm g} {}_{\gamma\delta} h^{\gamma\delta}
= \lie_\zeta g^{\alpha\beta}.
\label{eq:gaugeinvh3}
\end{gather}
\end{subequations}
Upon a gauge transformation \eq{eq:gauge}, one obtains
\begin{subequations}
\begin{gather}
h'^{\alpha\beta} = \invh'^{\alpha\beta} + \ninvh'^{\alpha\beta},
\\
\invh'^{\alpha\beta}
= \oper{\Pi}^{\alpha\beta}_{\rm inv} {}_{\gamma\delta}h'^{\gamma\delta}
= \invh^{\alpha\beta} + \oper{\Pi}^{\alpha\beta}_{\rm inv} {}_{\gamma\delta} \lie_\xi g^{\alpha\beta}
= \invh^{\alpha\beta},
\label{eq:psi}
\\
\ninvh'^{\alpha\beta}
= \oper{\Pi}^{\alpha\beta}_{\rm g} {}_{\gamma\delta}h'^{\gamma\delta}
= \ninvh^{\alpha\beta} + \lie_\xi g^{\alpha\beta}
= \lie_{\zeta +\xi} g^{\alpha\beta},
\label{eq:phi}
\end{gather}
\end{subequations}
where we used \Eqs{eq:proj0}--\eq{eq:PiLie2}. This means that $\smash{\ninvh^{\alpha\beta}}$, which is encoded by the four functions $\zeta^\mu$, does not contain gauge-independent information. Hence, any solution that has nonzero $\smash{\ninvh^{\alpha\beta}}$ and zero $\smash{\invh^{\alpha\beta}}$ can be classified as a coordinate artifact. In contrast, $\smash{\invh^{\alpha\beta}}$ is gauge-invariant by \Eq{eq:psi}. By the argument presented in \Sec{sec:intro}, it is encoded by six independent functions, or gauge-invariant degrees of freedom. Also note that $\smash{\invh^{\alpha\beta}}$ does \textit{not} necessarily satisfy the Lorenz-gauge condition $\smash{\del_\beta \invh^{\alpha\beta} = 0}$.

Finding an explicit formula for the Green's operator $\smash{\oper{\Xi}^\alpha{}_\beta}$ that determines $\smash{\invh^{\alpha\beta}}$ [\Eq{eq:gaugeinvh2}] for a specific background geometry is beyond the scope of this paper, because our primary concern is the general framework needed for a dispersive-GW theory. (This is similar to the approach taken by others; for example, see \Refs{ref:nakamura13, ref:frob18c, ref:frob18d}.) For a smooth background metric, $\smash{\oper{\Xi}^\alpha{}_\beta}$ can be found asymptotically within any predefined accuracy using methods of the Weyl symbol calculus (\App{app:Xi}). The proof of existence of the exact operator and an asymptotic approximation of its Weyl symbol is, as usual \cite{book:tracy, my:ql}, sufficient within fundamental wave theory to properly define waves as dynamic objects. We elaborate on this subject in application to GWs in \Ref{tex:mygwquasi}. Alternatively, recursive construction of a parametrix can be used for small distances. This method can yield a converging expansion of the Green's operator even for arbitrary globally hyperbolic spacetimes \cite{tex:baer08}. An example using Hadamard parametrices can be found in \Ref{ref:frob18b}. Also note that the very expression \eq{eq:gaugeinvh2} for the invariant part of the metric perturbation is not unique in general. In particular, any function of $\smash{\psi^{\alpha\beta}}$ is also an invariant.

\subsection{Six gauge invariants}

The six independent functions still need to be extracted from the sixteen gauge-invariant functions $\invh^{\alpha\beta}$. To do so, let us consider $h^{\alpha\beta}$ as a 16-dimensional (16-D) field $h^a$, or $\vec{h}$ in the index-free notation, of the form
\begin{gather}
\vec{h} = (h^{00}, h^{01}, h^{02}, h^{03}, h^{10}, \ldots, h^{32}, h^{33})^\intercal,
\label{eq:perturbvec}
\end{gather}
where $^\intercal$ denotes transpose. In other words,
\begin{align}
h^a = h^{\alpha\beta}, & \quad h_b = h_{\gamma\delta},\\
\{\alpha,\beta\} = \iota(a), & \quad \{\gamma,\delta\} = \iota(b),\label{eq:iot0}
\end{align}
where the index function $\iota$ is defined via
\begin{gather}
\iota(a) \doteq \big\{
1 + \lfloor (a-1)/4 \rfloor,
1 + (a-1)\,\text{mod}\,4
\big\}.
\end{gather}
(Here and further, Latin indices from the beginning of the alphabet range from 1 to 16.) Let us define $\mcu{H}_1$ as a Hilbert space of one-component functions on the background spacetime with the usual inner product $\braket{\cdot\,, \cdot}_1$. Then, the 16-D fields~\eq{eq:perturbvec} can be considered as vectors in the Hilbert space $\mcu{H}_{16}$ that is the tensor product of 16 copies of $\mcu{H}_1$, with the inner product
\begin{gather}
\braket{\vec{\xi}, \vec{\varphi}} = \int \dd^4 x\sqrt{-g}\, \xi_a^*\, \varphi^a 
= \sum_{a=1}^{16} \braket{\xi_a, \varphi^a}_1,
\end{gather}
where $g \doteq \det g_{\alpha\beta}$. (Unlike in the rest of the paper, summation is shown explicitly here in order to emphasize the difference between $\braket{\cdot\,, \cdot}$ and $\braket{\cdot\,, \cdot}_1$.) Then, $\smash{\oper{\Pi}_{\rm inv}^{\alpha\beta}{}_{\gamma\delta}}$ induces an operator $\smash{\oper{\Pi}^a{}_b}$ on $\mcu{H}_{16}$ defined via
\begin{gather}\label{eq:Pidef}
\oper{\Pi}^a{}_b h^b \doteq \oper{\Pi}_{\rm inv}^{\alpha\beta}{}_{\gamma\delta} h^{\gamma\delta},
\end{gather}
where we again assumed the notation as in \Eq{eq:iot0}. From \Eqs{eq:props}, one finds that
\begin{subequations}
\begin{gather}\label{eq:propsv}
\oper{\Pi}^a{}_b
\oper{\Pi}^b{}_c 
= \oper{\Pi}^a {}_c,
\\
\oper{\Pi}^a {}_b \lie_u g^b = 0.
\end{gather}
\end{subequations}

Equation \eq{eq:propsv}, which in the index-free notation can be written as $\boper{\Pi}^2 = \boper{\Pi}$, means that $\boper{\Pi}$ is a projector. (Note that $\boper{\Pi}^\dag \ne \boper{\Pi}$, so the projector is not orthogonal but oblique.) Hence, each eigenvalue of $\boper{\Pi}$ is either zero or unity and $\boper{\Pi}$ is diagonalizable. This means that $\boper{\Pi}$ can be represented~as
\begin{gather}\label{eq:PiJ}
\boper{\Pi} = \boper{\mT}\boper{J}\boper{\mT}^{-1},
\end{gather}
where $\smash{\boper{\mT}}$ is a diagonalizing transformation and the operator $\boper{J}$ is such that each component of the vector $\boper{J}\vec{h}$ equals either zero or the corresponding component of $\vec{h}$ for any~$\vec{h}$. Like $\smash{\oper{\Xi}^\alpha{}_\beta}$, the diagonalizing transformation cannot be found exactly but can be found asymptotically using methods of the Weyl symbol calculus if the inhomogeneity of the background metric is weak. In this sense, our identification of the gauge invariants is intended as an algorithm rather than as an explicit answer.

Each linear operator in $\mcu{H}_{16}$ is a $16 \times 16$ matrix of operators on $\mcu{H}_1$. Then, $\boper{J}$ must be represented by a constant matrix $\matr{J}$ of the form
\begin{gather}
\matr{J} = \text{diag}\,\big\{
\underbrace{1,1, \ldots ,1}_n,\underbrace{0,0,\ldots,0,0}_{16-n}
\big\},
\label{eq:J}
\end{gather}
where, for clarity, we have ordered the basis such that the nonzero eigenvalues are grouped together and have indices \(1,\ldots,n\). The gauge-invariant part of $\vec{h}$, which is given by \Eq{eq:gaugeinvh2}, can now be expressed as $\vec{\psi} = \boper{\Pi}\vec{h}$. Using \Eq{eq:PiJ}, one can also rewrite this as
\begin{gather}\label{eq:Psi}
\vec{\psi} = \boper{\mT} \vec{\Psi},
\quad
\vec{\Psi} = \matr{J} \boper{\mT}^{-1} \vec{h}.
\end{gather}
Because $\vec{h}$ is an arbitrary vector field parameterized by 16 functions and $\boper{\mT}$ is invertible, the field $\boper{\mT}^{-1} \vec{h}$ is also parameterized by 16 functions. Then, $\vec{\Psi}$ is parameterized by $n$ functions. But we know that $\vec{\psi}$ is parameterized by 6 functions (\Sec{sec:intro}), and thus so is $\vec{\Psi}$. Then, $n = 6$, and the nonzero elements of $\vec{\Psi}$ are the sought invariants.

In summary, to find the gauge invariants, one needs to find the diagonalizing transformation $\smash{\oper{\mT}^a{}_b}$ that brings $\smash{\oper{\Pi}^a{}_b}$ to the form given by \Eqs{eq:PiJ} and \eq{eq:J}. Then, the invariants can be found as
\begin{gather}\label{eq:Psi2}
\Psi^s = J^s{}_b (\oper{\mT}^{-1})^b{}_c h^c, 
\quad
s = 1, 2, \ldots, 6.
\end{gather}

\section{Example: Minkowski background}
\label{sec:eucl}

Except for toy models, problems involving wave propagation through inhomogeneous matter have no generic symmetries, so case studies are of little interest within the scope of this paper. What matter instead are the existence theorems, local analysis, and asymptotics \cite{book:tracy}. Hence, for an example, we will discuss only the simplest solvable case here, specifically, the case of the Minkowski background. Although interactions with matter generally curve the background, the Minkowski-background model can be a valid approximation for gravitational modes with a high refraction index \cite{my:gwinvar}. In addition, this example is instructive in that it allows direct benchmarking of our framework against known results.

\subsection{Gauge invariants}
\label{sec:gi}

In vacuum, when $R^\alpha{}_\beta \to 0$, one has $\smash{\oper{\Xi}^\alpha{}_\beta \to -\delta^\alpha_\beta \del^{-2}}$. For the Minkowski background, $\smash{\oper{\Xi}^\alpha{}_\beta}$ is further simplified~to  
\begin{gather}\label{eq:Xipd}
\oper{\Xi}^\alpha{}_\beta \to -\delta^\alpha_\beta \pd^{-2}.
\end{gather}
Here, $\smash{\pd^{-2}}$ is the operator inverse to $\smash{\pd^2 \doteq \pd_\mu \pd^\mu}$; \ie $\smash{\varphi^\alpha} = \smash{\pd^{-2} q^\alpha}$ is the solution of $\smash{\pd^2 \varphi^\alpha = q^\alpha}$ (\App{app:Xi}). Formally, $\smash{\pd^{-2}}$ is singular on free vacuum GWs, but the vacuum case can still be considered as a limit (\Sec{sec:vacuumgw}).

Using \Eq{eq:Xipd}, one can rewrite \Eqs{eq:PiPi2} as
\begin{gather}\label{eq:Piaux}
\oper{\Pi}^{\alpha\beta}_{\rm inv} {}_{\gamma\delta} 
=
\delta^\alpha_{(\gamma}\delta^\beta_{\delta)}
-2 \, \pd^{-2} \pd^{(\alpha} \delta^{\beta)}_{(\gamma} \pd^{\phantom{\beta)}}_{\delta)}
+ \pd^{-2} \pd^\alpha \pd^\beta g_{\gamma\delta}.
\end{gather}
Let us consider this operator in the Fourier representation, in which case it becomes a local matrix function of the wavevector $k_\mu$; namely, $\smash{\oper{\Pi}^{\alpha\beta}_{\rm inv} {}_{\gamma\delta}} = \Pi^{\alpha\beta}_{\rm inv} {}_{\gamma\delta}$,
\begin{gather}
\Pi^{\alpha\beta}_{\rm inv} {}_{\gamma\delta}
= \delta^\alpha_{(\gamma}\delta^\beta_{\delta)} 
- \frac{2k^{(\alpha}_{\phantom{\beta)}} \delta^{\beta)}_{(\gamma}k^{\phantom{\beta)}}_{\delta)}}{k^2}
+ g_{\gamma \delta}\,\frac{k^\alpha k^\beta}{k^2}.
\label{eq:PiMink}
\end{gather}
Using that $\del_\mu \to \pd_\mu \to \ii k_\mu$ in the Fourier representation [and in particular, \(\lie_u g^{\alpha\beta} = -2\ii k^{(\alpha} u^{\beta)}\)], the properties~\eq{eq:props} are easily verified. (At $k^2 = 0$, the usual rules of resonant-pole manipulation apply \cite{ref:frob18b}, but for the discussion below, which is restricted to the spectral representation, these details are not important.) One also finds by a direct calculation \cite{foot:math} that, as expected from \Eqs{eq:PiJ} and \eq{eq:J},
\begin{gather}
\text{rank}\,\matr{\Pi} = 6.
\end{gather}

The invariant part of the metric perturbation \eq{eq:gaugeinvh2} is now given by $\smash{\invh^{\alpha\beta}} = \smash{\Pi^{\alpha\beta}_{\rm inv} {}_{\gamma\delta} h^{\gamma\delta}}$, or explicitly,
\begin{gather}
\invh^{\alpha\beta}
= h^{\alpha\beta}
- \frac{k^\alpha k_\mu}{k^2}h^{\mu\beta}
- \frac{k^\beta k_\mu}{k^2}h^{\alpha\mu}
+ \frac{k^\alpha k^\beta}{k^2} h,
\end{gather}
where $h \doteq \smash{\text{tr}\,h^{\alpha\beta}}$. Without loss of generality, let us assume coordinates such that
\begin{gather}
k^\alpha = (\omega,0,0,\spatialk),
\label{eq:kz}
\end{gather}
where $\spatialk$ is the spatial wavenumber. Using this, the fact that $k^2 = \spatialk^2 - \omega^2$, and also \Eq{eq:iot0}, the 16-D vector \(\vec{\psi}\) is found to be:
\begin{gather}\notag
\vec{\psi} = \frac{1}{k^2}
\begin{pmatrix}
 {h^{00} \spatialk^2 -2 h^{03}\omega\spatialk + \omega^2(h^{11} + h^{22} + h^{33} )}
  \smallskip\\
 {h^{01} \spatialk^2-h^{13} \omega \spatialk } 
 \smallskip\\
 {h^{02} \spatialk^2-h^{23} \omega \spatialk } \smallskip\\
 {(h^{11}+h^{22}) \spatialk \omega } 
\smallskip\\
{h^{01} \spatialk^2-h^{13} \omega \spatialk } 
 \smallskip\\
 h^{11} ({\spatialk^2-\omega^2})
\smallskip\\
 h^{12}({\spatialk^2-\omega^2})
\smallskip\\
 {h^{01} \omega \spatialk-h^{13} \omega^2}
\smallskip\\
{h^{02} \spatialk^2-h^{23} \omega \spatialk } \smallskip\\
 h^{12}({\spatialk^2-\omega^2})
\smallskip\\
 h^{22} ({\spatialk^2-\omega^2})
\smallskip\\
 {h^{02} \omega \spatialk-h^{23} \omega^2}
\smallskip\\
 {(h^{11}+h^{22}) \spatialk \omega } 
\smallskip\\
 {h^{01} \omega \spatialk-h^{13} \omega^2}
\smallskip\\
 {h^{02} \omega \spatialk-h^{23} \omega^2}
\smallskip\\
 \spatialk^2 (-h^{00} + h^{11} +h^{22}) + 2 h^{03}\omega\spatialk - h^{33} \omega^2
\end{pmatrix}.
\end{gather}

In order to extract the six gauge invariants from this $\vec{\psi}$, notice that the operator \eq{eq:Pidef} is represented by a local function of $k_\mu$, $\boper{\Pi} = \matr{\Pi}$, and thus so is the diagonalizing transformation \eq{eq:PiJ}. Specifically, $\smash{\boper{\mT}} = \matr{\mT}$, and the columns of the matrix $\matr{\mT}$ are just the eigenvectors of $\matr{\Pi}$:
\begin{gather}\label{eq:eiv}
\vec{\mT} = (\vec{v}_1 \kern 5pt \vec{v}_2 \kern 5pt \ldots \kern 5pt \vec{v}_{16}),
\quad
\matr{\Pi}\vec{v}_a = \lambda_a \vec{v}_a,
\end{gather}
where $\lambda_a \in \{0, 1\}$. The calculation of these eigenvectors and of the matrix $\smash{\matr{\mT}^{-1}}$ can be automated \cite{foot:math}, and the six gauge invariants \eq{eq:Psi2} are readily found to be
\begin{gather}
\vec{\Psi} = 
\begin{pmatrix}
 \displaystyle
 \frac{\spatialk^2(-h^{00}+h^{11}+h^{22}) + 2 \omega\spatialk h^{03} - \omega^2 h^{33}}{\spatialk^2-\omega^2}
 \\[10pt]
 \displaystyle
 \frac{\omega\spatialk h^{01} - \omega^2 h^{13}}{\spatialk^2 - \omega^2}
\\[10pt]
 \displaystyle
 \frac{\omega\spatialk h^{02} - \omega^2 h^{23}}{\spatialk^2 - \omega^2}
\\[10pt]
 \displaystyle
 \frac{\omega\spatialk(h^{11} + h^{22})} {\spatialk^2 - \omega^2}
 \\[10pt]
 h^{22}
 \\[5pt]
 \displaystyle
 h^{12}
\end{pmatrix}.
\label{eq:q6}
\end{gather}
The coordinate representation of these invariants is found by taking the inverse Fourier transform of \Eq{eq:q6}. 

Our result is in agreement with Eqs.~(2.45)--(2.47) in \Ref{ref:flanagan05} (which operates with $h_{\alpha\beta}$ instead of our $h^{\alpha\beta}$). This is seen from the fact that any linear combinations of our $\Psi^s$ are gauge invariants too. In other words, instead of $\smash{\Psi^s}$, one can introduce the invariants as $\smash{\bar{\Psi}^s}$ given by
\begin{gather}\label{eq:mcI}
\bar{\Psi}^s \doteq C^s{}_r \Psi^r, 
\quad r, s = 1, 2, \ldots , 6,
\end{gather}
or $\bar{\vec{\Psi}} = \vec{C}\vec{\Psi}$ in the index-free representation, where $\vec{C}$ is an arbitrary matrix that may depend on $k_\mu$. This is particularly convenient at \(\smash{k^2} \equiv \smash{\spatialk^2 - \omega^2} \to 0\), when $\vec{\Psi}$ becomes singular. Specifically, by choosing
\begin{gather}
\vec{C} = \text{diag}\,\big\{k^2, k^2, k^2, k^2, 1, 1\big\},
\end{gather}
we obtain invariants that are well-behaved at all~$k_\mu$:
\begin{gather}\label{eq:bPsi}
\bar{\vec{\Psi}} = \begin{pmatrix}
 \displaystyle
 \spatialk^2(-h^{00}+h^{11}+h^{22}) + 2 \omega\spatialk h^{03} - \omega^2 h^{33}
 \smallskip\\
 \displaystyle
 \omega\spatialk h^{01} - \omega^2 h^{13}
\smallskip\\
 \displaystyle
 \omega\spatialk h^{02} - \omega^2 h^{23}
\smallskip\\
 \displaystyle
 \omega\spatialk(h^{11} + h^{22})
 \smallskip\\
 h^{22}
 \smallskip\\
 \displaystyle
 h^{12}
\end{pmatrix}.
\end{gather}
As also mentioned in \Sec{sec:gen}, these invariants are not unique in that any function of them is an invariant too.

Let us also discuss \textit{why} the original vectors $\vec{\psi}$ and $\vec{\Psi}$ are singular at \(\smash{k^2} \to 0\). In this limit, the vectors $\vec{v}_a$ [\Eq{eq:eiv}] are well-behaved, and thus so is the matrix~$\matr{\mT}$. However, they cease to be linearly independent at $k^2 = 0$, so $\smash{\matr{\mT}^{-1}}$ becomes singular, and as a result, $\vec{\Pi}$ becomes singular too. This means that no finite invariant projection of a generic $\smash{h^{\alpha\beta}}$ can be defined in the Fourier space at $k^2 = 0$. The corresponding gauge-dependent part \(\phi^{\alpha\beta}\) becomes singular as well in this limit, as seen from \Eqs{eq:zeta} and \eq{eq:gaugeinvh3}, where $\smash{\oper{\Xi}^\alpha{}_\beta}$ becomes singular (\App{app:Xi}).\footnote{This is the same effect as the unlimited growth, at $x^\mu \to \infty$, of the gauge field that brings a generic $\smash{h^{\alpha\beta}}$ to the Lorenz gauge. See \App{app:Xi} in conjunction with \Eq{eq:xisol}, which is commonly known for the Minkowski background \cite{foot:schutz}.} Still, our general formulation correctly predicts the invariants \eq{eq:bPsi} at zero $k^2$, and these invariants can be related to vacuum GWs as discussed in the next section.

\subsection{Free GWs in the Minkowski space}
\label{sec:vacuumgw}

By comparing \Eq{eq:Piaux} with, for example, Eqs.~(5.4) and~(2.7) in \Ref{ref:isaacson68a}, one finds that the equation for vacuum GWs in the Minkowski spacetime can be expressed as
\begin{gather}\label{eq:vw}
\oper{D}^{\alpha\beta}{}_{\gamma\delta} h^{\gamma\delta} = 0,
\quad 
\oper{D}^{\alpha\beta}{}_{\gamma\delta} = \pd^2 \oper{\Pi}^{\alpha\beta}_{\rm inv} {}_{\gamma\delta}.
\end{gather}
In other words, in the special case of the Minkowski spacetime, the dispersion operator \(\smash{\oper{D}^{\alpha\beta}{}_{\gamma\delta}}\) of vacuum GWs is exactly $\pd^2$ times the operator that projects a metric perturbation on the invariant subspace. Thus, as expected, using the operators introduced in this paper, the wave equation for the GWs in vacuum can be shown to directly specify the gauge invariants and naturally weed out the gauge artifacts.

Let us also briefly discuss monochromatic waves,\footnote{Cf.\ a similar discussion in \Ref{ref:maccallum73}, except their Eq.~(3.6) describes the \textit{trace-reversed} metric perturbation.} in which case, \Eq{eq:vw} becomes
\begin{gather}\label{eq:gweq}
k^2 \,\Pi^{\alpha\beta}_{\rm inv} {}_{\gamma\delta}
\, h^{\gamma\delta}
= 0,
\end{gather}
where the matrix $\smash{k^2 \,\Pi^{\alpha\beta}_{\rm inv}}$ is well-behaved for all $k_\mu$. Equation \eq{eq:gweq} can be written as the following six of equations, which determine the six gauge invariants \eq{eq:bPsi}:
\begin{subequations}\label{eq:hvw}
\begin{gather}
 {\spatialk^2 h^{00} +\omega(-2\spatialk h^{03} + \omega h^{33} )} = 0, \label{eq:hvw1}\\
 {\spatialk^2 h^{01}- \omega \spatialk h^{13} } = 0, \label{eq:hvw2}\\
 {\spatialk^2 h^{02}- \omega \spatialk h^{23}} = 0, \label{eq:hvw3}\\ 
 \spatialk \omega {(h^{11}+h^{22})} = 0 ,\label{eq:hvw4}\\
 k^2 (h^{11}-h^{22}) = 0, \label{eq:hvw5}\\
 k^2 h^{12} = 0. \label{eq:hvw6}
\end{gather}
\end{subequations}
For \(k^2 \ne 0\), \Eqs{eq:hvw} indicate that all the six invariants \eq{eq:bPsi} are zero, so only coordinate waves are possible in this case. For \(k^2=0\), \Eqs{eq:hvw1}--\eq{eq:hvw4} yield
\begin{gather}
\bar{\Psi}^1 = \bar{\Psi}^2 = \bar{\Psi}^3 = \bar{\Psi}^4 = 0,
\end{gather}
and in particular, $h^{11}+h^{22} = 0$. However, \Eqs{eq:hvw5} and \eq{eq:hvw6} are satisfied identically at $k^2 = 0$, so the other two invariants,
\begin{gather}
\bar{\Psi}^5 = h^{22} = - h^{11},
\quad
\bar{\Psi}^6 = h^{12} = h^{21},
\end{gather}
can be arbitrary and represent the two tensor modes of the GWs in vacuum \cite{ref:flanagan05}.

\section{Conclusions}
\label{sec:conclusion}

In summary, we propose a method for identifying the gauge-invariant part $\smash{\invh^{\alpha\beta}}$ of the metric perturbation $\smash{h^{\alpha\beta}}$ within linearized gravity for an arbitrary background metric $\smash{g_{\alpha\beta}}$ assuming that the inverse of a hyperbolic operator \(\smash{\oper{Q}^\alpha{}_\beta}\) \eq{eq:QX}. Specifically, we show that $\smash{\invh^{\alpha\beta}} = \smash{\oper{\Pi}^{\alpha\beta}_{\rm inv} {}_{\gamma\delta}h^{\gamma\delta}}$, where $\smash{\oper{\Pi}^{\alpha\beta}_{\rm inv} {}_{\gamma\delta}}$ is a linear operator given by \Eq{eq:PiPi2}. The six independent functions from the sixteen gauge-invariant functions $\invh^{\alpha\beta}$ can be found using \Eq{eq:Psi2}. These results lead to a gauge-invariant quasilinear theory of dispersive gravitational waves in an arbitrary background, as discussed in a companion paper \cite{tex:mygwquasi} (see also \Ref{my:ql}). For the Minkowski background, the well-known dispersion operator of linear GWs in vacuum is proportional to $\smash{\oper{\Pi}^{\alpha\beta}_{\rm inv}{}_{\gamma\delta}}$ [\Eq{eq:vw}], and thus specifies the gauge invariants directly streamlining the process of removing the gauge artifacts. We also show that this general formulation systematically yields the six known gauge invariants for the Minkowski background.

This material is based upon the work supported by National Science Foundation under the grant No. PHY~1903130.

\appendix

\section{Asymptotic representation of $\boldsymbol{\oper{\Xi}^{\alpha}{}_\beta}$}
\label{app:Xi}

An asymptotic approximation for the Green's operator $\smash{\oper{\Xi}^{\alpha}{}_{\beta}}$ as the inverse of $\smash{\oper{Q}^\alpha{}_\beta}$ [\Eq{eq:QX}] can be constructed using methods of the Weyl symbol calculus. These methods may not be particularly popular in general relativity, but they have become \textit{de~facto} standard in fundamental wave theory \cite{book:tracy} that shifts focus from specific wave equations to a more generic description applicable to waves in any dispersive medium. A systematic application of the Weyl symbol calculus involves mapping an operator to a function called the Weyl symbol,\footnote{For a homogeneous medium, the Weyl symbol of a given operator is obtained by replacing each $\pd_\alpha$ with $-\ii k_\alpha$. In an inhomogeneous medium, the procedure is more elaborate \cite{book:tracy, my:quasiop1}.} approximating this symbol, and then mapping the result back to the operator space \cite{ref:mcdonald88, my:quasiop1, my:ql}. Finding Green's operators by inverting symbols of the dispersion operators is a common practice as well; for example, see \Refs{my:ql, my:qponder} for a modern reformulation of classic results. These calculations can be done within any predefined accuracy for smooth backgrounds. However, doing this carefully requires introducing machinery that is beyond the scope of this paper, so here we opt for a less rigorous but more intuitive argument.

The operator $\smash{\oper{\Xi}^{\alpha}{}_{\beta}}$ defined in \Eq{eq:Xi} can be written in the index-free representation~as
\begin{gather}
\boper{\Xi} = - (\del^2 + \boper{R})^{-1},
\label{eq:Xi1}
\end{gather}
where $\del^2 \doteq \del_\mu \del^\mu$, $\boper{R}$ is the operator whose coordinate representation is the Ricci tensor $\smash{R^\alpha{}_\beta}$, and $^{-1}$ denotes the operator inverse. In order for this inverse to exist (approximately), we assume the adiabatic limit. Specifically, we assume that the characteristic GW wavelength $\lambda$ is much smaller than the characteristic radius $L$ of the spacetime curvature, \ie when $\epsilon \doteq \lambda/L \ll 1$. Assuming the ordering $\lambda = \mc{O}(1)$ and $L = \mc{O}(\epsilon^{-1})$, one has $\del^2 = \mc{O}(1)$ and $\boper{R} = \mc{O}(\epsilon^2)$. Then, 
\begin{gather}
\boper{\Xi} = - \del^{-2} + \del^{-2} \boper{R}\, \del^{-2} + \mc{O}(\epsilon^4),
\label{eq:Xi2}
\end{gather}
where $\del^{-2}$ is the inverse of $\smash{\del^2}$; \ie $\smash{\varphi^\alpha} = \smash{\del^{-2} q^\alpha}$ is defined as the solution of $\smash{\del^2 \varphi^\alpha = q^\alpha}$. 

Because the operators in \Eqs{eq:Xi1} and \eq{eq:Xi2} are intended to act specifically on vector fields, one can also write them explicitly. For example, in normal coordinates, one has (\App{app:del2normal})
\begin{gather}
\label{eq:del2normal}
\del^2 = \pd^2 - \frac{\boper{R}}{3},
\end{gather}
and the corresponding inverse is
\begin{gather}
\del^{-2} = \pd^{-2} + \frac{1}{3}\,\pd^{-2} \boper{R}\, \pd^{-2} + \mc{O}(\epsilon^4),
\end{gather}
so \Eq{eq:Xi2} leads to
\begin{gather}
\boper{\Xi} = - \pd^{-2} + \frac{2}{3}\,\pd^{-2} \boper{R}\, \pd^{-2} + \mc{O}(\epsilon^4).
\label{eq:Xi3}
\end{gather}

The operator $\pd^{-2}$ that enters here is understood as the Green's operator of the equation
\begin{gather}\label{eq:phq}
\pd^2 \varphi^\alpha = q^\alpha.
\end{gather}
(This is the same equation that emerges in the well-known linear gravity in the Minkowski background \cite{foot:schutz}; see also \Eq{eq:xisol}.) Suppose that the right-hand side of \Eq{eq:phq} is quasimonochromatic, \ie $\smash{q^\alpha = Q^\alpha \exp[\ii \theta(x^\mu)]}$ with $\pd_\beta Q^\alpha = \mc{O}(\epsilon)$ and $\pd_\beta k_\alpha = \mc{O}(\epsilon)$, where $k_\alpha \doteq \pd_\alpha \theta$ is the local wavevector. Then, 
\begin{gather}\label{eq:keps}
\pd^{-2} = (k_\mu k^\mu)^{-1} + \oper{\Delta},
\end{gather}
where $\oper{\Delta} = \mc{O}(\epsilon)$ is a differential operator to act on the envelope $\smash{Q^\alpha}$. If $k^2 \doteq k_\mu k^\mu$ approaches zero, as would be the case for GWs in the Minkowski vacuum, then $\varphi^\alpha$ grows indefinitely at $x^\mu \to \infty$. This is due to the fact that at $k^2 \to 0$, $q^\alpha$ acts as a resonant driving force for $\varphi^\alpha$. No quasimonochromatic solution is possible in this case, and $\varphi^\alpha$ necessarily diverges at infinity. In particular, this means that even if the Fourier spectrum of $q^\alpha$ is analytic but includes harmonics with $k^2 = 0$, the Fourier spectrum of the corresponding $\varphi^\alpha$ is singular. 

This indicates that the case $k^2 = 0$ cannot be treated within the adiabatic approximation that we assume in this paper. However, it still can be considered as a limit, as discussed in \Sec{sec:eucl}. Also, no such issues arise in problems that involve GW--matter coupling, because then $k^2 \ne 0$. In this case, the term $\oper{\Delta}$ in \Eq{eq:keps} can be calculated too, but there is no need to do this explicitly in the present paper. (The general approach to such calculations is described, for example, in \Ref{my:quasiop1}.) What matters instead is that $\smash{\boper{\Xi}}$ is a well-defined object that, in principle, can be found within any predefined accuracy. As usual in fundamental wave theory \cite{book:tracy}, the zeroth-order or first-order approximation of the Green's operator often suffices for practical applications \cite{my:quasiop1, my:qponder, my:ql, my:wkeadv}.

\section{Derivation of \Eqs{eq:PLie0} and~\eq{eq:PLie}}
\label{app:PLie}

Using \Eq{eq:P0} for \(\smash{\oper{\pi}^{\alpha\beta}{}_{\gamma\delta}}\) and \Eq{eq:lie} for $\smash{\lie_u g^{\gamma\delta}}$, one obtains
\begin{align}
\oper{\pi}^{\alpha\beta}{}_{\gamma\delta}
\lie_u g^{\gamma\delta} 
&=
- \big(
\delta_\gamma^\alpha \delta_\delta^\beta
+ \del^\alpha {\oper{\Xi}^\beta} {}_\gamma \del_\delta
\nonumber
\\&
\qquad + \del^\beta {\oper{\Xi}^\alpha} {}_\gamma \del_\delta \big)
\left(\del^\gamma u^\delta
+ \del^\delta u^\gamma \right).
\end{align}
Then using \Eq{eq:Ru} in the above equation yields
\begin{align}
\oper{\pi}^{\alpha\beta}{}_{\gamma\delta}
\lie_u g^{\gamma\delta} 
&= - 2 \del^{(\alpha} u^{\beta)}
- 2 \del^{(\alpha} {\oper{\Xi}^{\beta)}} {}_\gamma
\left[
\del_\delta,\del^\gamma \right] u^\delta
\nonumber
\\&
\qquad - 2 \del^{(\alpha} {\oper{\Xi}^{\beta)}} {}_\gamma
\del^\gamma \del_\delta u^\delta
- \del^{(\alpha} {\oper{\Xi}^{\beta)}} {}_\gamma
\del^2 u^\gamma
\nonumber
\\&
= -2 \del^{(\alpha} u^{\beta)}
- 2 \del^{(\alpha} {\oper{\Xi}^{\beta)}} {}_\gamma
\left(
\delta_\delta^\gamma
\del^2 + {R^\gamma}_\delta\right) u^\delta
\nonumber
\\&
\qquad - 2 \del^{(\alpha} {\oper{\Xi}^{\beta)}} {}_\gamma
\del^\gamma \del_\delta u^\delta.
\end{align}
Using \Eq{eq:QX} for \(\smash{\oper{Q}^\alpha{}_\beta}\) in combination with \Eq{eq:Xi}, one obtains
\begin{align}
\oper{\pi}^{\alpha\beta}{}_{\gamma\delta}
\lie_u g^{\gamma\delta} 
&= -2 \del^{(\alpha} u^{\beta)}
+ 2 \del^{(\alpha} {\oper{\Xi}^{\beta)}} {}_\mu
{\oper{Q}^\mu} {}_\delta u^\delta
\nonumber
\\&
\qquad - 2 \del^{(\alpha} {\oper{\Xi}^{\beta)}} {}_\mu
\del^\mu \del_\delta u^\delta
\nonumber
\\&
= -2 \del^{(\alpha} u^{\beta)}
+ 2 \del^{(\alpha}_{\phantom{\delta}} \delta^{\beta)}_\delta u^\delta
\nonumber
\\&
\qquad - 2 \del^{(\alpha} {\oper{\Xi}^{\beta)}} {}_\mu
\del^\mu \del_\delta u^\delta
\nonumber
\\&
= -2 \del^{(\alpha} {\oper{\Xi}^{\beta)}} {}_\mu
\del^\mu \del_\delta u^\delta.
\label{eq:aux21}
\end{align}
For $\del_\delta u^\delta = 0$, this leads to $\smash{\oper{\pi}^{\alpha\beta}{}_{\gamma\delta}
\lie_u g^{\gamma\delta}} = 0$, which is \Eq{eq:PLie0}. Otherwise, notice that
\begin{gather}
2\del_\delta u^\delta = 2g_{\delta\gamma} \del^\gamma u^\delta
= 2 g_{\gamma\delta} \del^{(\gamma} u^{\delta)}
= - g_{\gamma\delta}  \lie_u g^{\gamma\delta}.
\end{gather}
Then, one can rewrite \Eq{eq:aux21} as
\begin{gather}
\oper{\pi}^{\alpha\beta}{}_{\gamma\delta}
\lie_u g^{\gamma\delta} 
= \del^{(\alpha} {\oper{\Xi}^{\beta)}} {}_\mu
\del^\mu g_{\gamma\delta} \lie_u g^{\gamma\delta},
\end{gather}
which is precisely \Eq{eq:PLie}.

\section{Derivation of \Eq{eq:proj2}}
\label{app:proj}

Using \Eq{eq:Piparallel}, we get
\begin{multline}
\oper{\Pi}_{\rm g}^{\alpha\beta}{}_{\gamma\delta}
\oper{\Pi}_{\rm g}^{\gamma\delta}{}_{\lambda\varepsilon}
= 4\del^{(\alpha}\oper{\Xi}^{\beta)}{}_{(\gamma} \del_{\delta)} \del^{(\gamma} \oper{\Xi}^{\delta)}{}_{(\lambda} \del_{\varepsilon)}
\\
-2\del^{(\alpha} \oper{\Xi}^{\beta)}{}_{(\gamma} \del_{\delta)} \del^{(\gamma} \oper{\Xi}^{\delta)}{}_\nu \del^\nu g_{\lambda\varepsilon}
\\
-2 \del^{(\alpha} \oper{\Xi}^{\beta)}{}_\mu \del^\mu g_{\gamma\delta} \del^{(\gamma} \oper{\Xi}^{\delta)}{}_{(\lambda} \del_{\varepsilon)}
\\
+ \del^{(\alpha} \oper{\Xi}^{\beta)}{}_\mu \del^\mu g_{\gamma\delta} \del^{(\gamma} \oper{\Xi}^{\delta)}{}_\nu \del^\nu g_{\lambda\varepsilon}.
\end{multline}
Let us simplify the individual terms on the right-hand side separately. We start by expanding one pair of symmetrized indices to get
\begin{align}
& 4 \del^{(\alpha} \oper{\Xi}^{\beta)}{}_{(\gamma} \del_{\delta)} \del^{(\gamma} \oper{\Xi}^{\delta)}{}_{(\lambda} \del_{\varepsilon)}
\nonumber \\& \qquad
=2 \del^{(\alpha} \oper{\Xi}^{\beta)}{}_\gamma \del_\delta \del^\gamma \oper{\Xi}^\delta{}_{(\lambda}\del_{\varepsilon)}
+2 \del^{(\alpha} \oper{\Xi}^{\beta)}{}_\gamma \del^2 \oper{\Xi}^\gamma{}_{(\lambda} \del_{\varepsilon)}
\nonumber \\& \qquad 
= 2\del^{(\alpha} \oper{\Xi}^{\beta)}{}_\gamma \del^\gamma \del_\delta \oper{\Xi}^\delta{}_{(\lambda} \del_{\varepsilon)}
+ 2\del^{(\alpha} \oper{\Xi}^{\beta)}{}_\gamma \del^2 \oper{\Xi}^\gamma{}_{(\lambda} \del_{\varepsilon)}
\nonumber \\& \qquad \qquad
+ 2\del^{(\alpha} \oper{\Xi}^{\beta)}{}_\gamma \left[\del_\delta,\del^\gamma\right] \oper{\Xi}^{\delta}{}_{(\lambda} \del_{\varepsilon)}.
\end{align}
Recognizing that the operator would act on a rank-2 tensor \(\smash{h^{\lambda\varepsilon}}\), we can use \Eq{eq:Ru} for the commutator; hence,
\begin{multline}
4 \del^{(\alpha} \oper{\Xi}^{\beta)}{}_{(\gamma} \del_{\delta)} \del^{(\gamma} \oper{\Xi}^{\delta)}{}_{(\lambda} \del_{\varepsilon)}
= 2\del^{(\alpha} \oper{\Xi}^{\beta)}{}_\gamma \del^\gamma \del_\delta \oper{\Xi}^\delta{}_{(\lambda} \del_{\varepsilon)}
\\
+ 2\del^{(\alpha} \oper{\Xi}^{\beta)}{}_\gamma \left(
{R^\gamma}_\delta + \delta_\delta^\gamma
\del^2 \right) \oper{\Xi}^{\delta}{}_{(\lambda} \del_{\varepsilon)}.
\end{multline}
The terms in the parenthesis on the right-hand side of the above equation can be expressed through \(\smash{\oper{Q}^\alpha{}_\beta}\) [\Eq{eq:QX}], which is also the inverse of \(\smash{\oper{\Xi}^\alpha{}_\beta}\) [\Eq{eq:Xi}]; hence,
\begin{align}
& 4 \del^{(\alpha} \oper{\Xi}^{\beta)}{}_{(\gamma} \del_{\delta)} \del^{(\gamma} \oper{\Xi}^{\delta)}{}_{(\lambda} \del_{\varepsilon)}
\nonumber \\& \quad
= 2\del^{(\alpha} \oper{\Xi}^{\beta)}{}_\gamma \del^\gamma \del_\delta \oper{\Xi}^\delta{}_{(\lambda} \del_{\varepsilon)}
- 2\del^{(\alpha} \oper{\Xi}^{\beta)}{}_\gamma  {\oper{Q}^\gamma} {}_\delta \oper{\Xi}^{\delta}{}_{(\lambda} \del_{\varepsilon)}
\nonumber \\& \quad
= 2\del^{(\alpha} \oper{\Xi}^{\beta)}{}_\gamma \del^\gamma \del_\delta \oper{\Xi}^\delta{}_{(\lambda} \del_{\varepsilon)}
- 2\del^{(\alpha} \oper{\Xi}^{\beta)}{}_{(\lambda} \del_{\varepsilon)}.
\end{align}
Using a similar process, the second term is found to be
\begin{align}
& 2\del^{(\alpha} \oper{\Xi}^{\beta)}{}_{(\gamma} \del_{\delta)} \del^{(\gamma} \oper{\Xi}^{\delta)}{}_\nu \del^\nu g_{\lambda\varepsilon}
\nonumber \\& \quad
=\del^{(\alpha} \oper{\Xi}^{\beta)}{}_\gamma \del_\delta \del^\gamma {\oper{\Xi}^\delta} {}_{\nu} \del^\nu g_{\lambda\varepsilon} +\del^{(\alpha} \oper{\Xi}^{\beta)}{}_\gamma \del^2 {\oper{\Xi}^\gamma} {}_\nu \del^\nu g_{\lambda\varepsilon}
\nonumber \\& \quad
=\del^{(\alpha} {\oper{\Xi}^{\beta)}}{}_\gamma
\left(
{R^\gamma}_\delta +\delta_\delta^\gamma \del^2\right) {\oper{\Xi}^\delta}{}_\nu \del^\nu g_{\lambda\varepsilon}
\nonumber \\& \qquad
+\del^{(\alpha} {\oper{\Xi}^{\beta)}}{}_\gamma \del^\gamma \del_\delta {\oper{\Xi}^\delta}{}_\nu \del^\nu g_{\lambda\varepsilon}
\nonumber \\& \quad
=-\del^{(\alpha} \oper{\Xi}^{\beta)}{}_\mu \del^\mu g_{\lambda\varepsilon}
+\del^{(\alpha} {\oper{\Xi}^{\beta)}}{}_\gamma \del^\gamma \del_\delta {\oper{\Xi}^\delta}{}_\nu \del^\nu g_{\lambda\varepsilon}.
\end{align}
The third and the fourth terms are simply
\begin{gather}
2 \del^{(\alpha} \oper{\Xi}^{\beta)}{}_\mu \del^\mu g_{\gamma\delta} \del^{(\gamma} \oper{\Xi}^{\delta)}{}_{(\lambda} \del_{\varepsilon)}
= 2 \del^{(\alpha} \oper{\Xi}^{\beta)}{}_\mu \del^\mu \del_\delta \oper{\Xi}^\delta{}_{(\lambda} \del_{\varepsilon)},
\nonumber\\
\del^{(\alpha} \oper{\Xi}^{\beta)}{}_\mu \del^\mu g_{\gamma\delta} \del^{(\gamma} \oper{\Xi}^{\delta)}{}_\nu \del^\nu g_{\lambda\varepsilon}
=
\del^{(\alpha} \oper{\Xi}^{\beta)}{}_\mu \del^\mu \del_\delta \oper{\Xi}^\delta{}_\nu \del^\nu g_{\lambda\varepsilon}.
\nonumber
\end{gather}
Combining all these expressions, we get
\begin{align}
\oper{\Pi}_{\rm g}^{\alpha\beta}{}_{\gamma\delta}
\oper{\Pi}_{\rm g}^{\gamma\delta}{}_{\lambda\varepsilon}
& = 2\del^{(\alpha} \oper{\Xi}^{\beta)}{}_\gamma \del^\gamma \del_\delta \oper{\Xi}^\delta{}_{(\lambda} \del_{\varepsilon)}
\notag\\
& - 2\del^{(\alpha} \oper{\Xi}^{\beta)}{}_{(\lambda} \del_{\varepsilon)}
+\del^{(\alpha} \oper{\Xi}^{\beta)}{}_\mu \del^\mu g_{\lambda\varepsilon}
\notag\\
& -\del^{(\alpha} {\oper{\Xi}^{\beta)}}_\gamma \del^\gamma \del_\delta {\oper{\Xi}^\delta}{}_\nu \del^\nu g_{\lambda\varepsilon}
\notag\\
&-2 \del^{(\alpha} \oper{\Xi}^{\beta)}{}_\mu \del^\mu \del_\delta \oper{\Xi}^\delta{}_{(\lambda} \del_{\varepsilon)}
\notag\\
&+ \del^{(\alpha} \oper{\Xi}^{\beta)}{}_\mu \del^\mu \del_\delta \oper{\Xi}^\delta{}_\nu \del^\nu g_{\lambda\varepsilon}.
\end{align}
Canceling the first term on the right-hand side with the fifth term, and the fourth term with the sixth term, we arrive at
\begin{gather}
\oper{\Pi}_{\rm g}^{\alpha\beta}{}_{\gamma\delta}
\oper{\Pi}_{\rm g}^{\gamma\delta}{}_{\lambda\varepsilon}
= - 2\del^{(\alpha} \oper{\Xi}^{\beta)}{}_{(\lambda} \del_{\varepsilon)}
+\del^{(\alpha} \oper{\Xi}^{\beta)}{}_\mu \del^\mu g_{\lambda\varepsilon}.
\end{gather}
Upon comparison with \Eq{eq:Piparallel}, this leads to \Eq{eq:proj2}.

\section{Derivation of \Eq{eq:del2normal}}
\label{app:del2normal}

For any vector field \(u^\alpha\), one has
\begin{align}
\del^\beta\del_\beta u^\alpha
&=\del^\beta \left(\pd_\beta u^\alpha
+ \Gamma^\alpha_{\beta\lambda} u^\lambda \right)
\nonumber
\\
&=\pd^\beta \left(\pd_\beta u^\alpha
+ \Gamma^\alpha_{\beta\lambda} u^\lambda \right)
\nonumber
\\
&\quad + g^{\beta\gamma}
\Gamma^\alpha_{\gamma\rho} \left(\pd_\beta u^\rho
+ \Gamma^\rho_{\beta\lambda} u^\lambda \right)
\nonumber
\\
&\quad -g^{\beta\gamma} \Gamma^\rho_{\beta\gamma} \left(\pd_\rho u^\alpha
+ \Gamma^\alpha_{\rho\lambda} u^\lambda \right),
\end{align}
where $\Gamma^\alpha_{\beta\gamma}$ are the Christoffel symbols. In normal coordinates, the Christoffel symbols are zero, but their derivatives are not. This leads to
\begin{gather}\label{eq:aux1}
\del^2 u^\alpha
=\pd^2 u^\alpha
+ u^\lambda \pd^\beta \Gamma^\alpha_{\beta\lambda}.
\end{gather}
The derivatives of the Christoffel symbols can be expressed through the Riemann tensor ${R^\rho}_{\sigma\mu\nu}$ \cite{ref:brewin98}:
\begin{gather}
\pd_\nu \Gamma_{\mu\sigma}^\rho
= -\frac{1}{3}\left(
{R^\rho}_{\sigma\mu\nu}
+ {R^\rho}_{\mu\sigma\nu}\right).
\end{gather}
Using the well-known symmetries of the Riemann tensor and of the Ricci tensor $R_{\sigma\nu} \doteq {R^\rho}_{\sigma\rho\nu}$, one then finds that
\begin{multline}\notag
\pd^\beta \Gamma_{\beta\lambda}^\alpha
= -\frac{1}{3}\left(
{R^\alpha}_{\lambda\beta}{}^\beta
+ {R^\alpha}_{\beta\lambda}{}^\beta\right)
= -\frac{1}{3}\,{R^\alpha}_{\beta\lambda}{}^\beta
\\
= -\frac{1}{3}\,{R_\lambda}^{\beta\alpha}{}_\beta
= -\frac{1}{3}\,{R^\beta}{}_{\lambda\beta}{}^{\alpha}
= -\frac{1}{3}\,{R}_{\lambda}{}^\alpha =  -\frac{1}{3}\,{R^\alpha}_{\lambda}.
\end{multline}
Hence, one can rewrite \Eq{eq:aux1}~as
\begin{gather}
\del^2 u^\alpha
=\pd^2 u^\alpha
-\frac{1}{3}\, {R^\alpha}_\beta u^\beta,
\end{gather}
or equivalently, as
\begin{gather}
(\del^2)^\alpha{}_\beta = \delta^\alpha_\beta \pd^2 - \frac{1}{3}\,R^\alpha{}_\beta.
\end{gather}
In the index-free representation, this leads to \Eq{eq:del2normal}.


\end{document}